\documentclass[conference]{IEEEtran}
\usepackage{cite}
\usepackage{amsmath,amssymb,amsfonts}
\usepackage{algorithmic}
\usepackage{braket}
\usepackage{graphicx}
\usepackage{textcomp}
\usepackage{subcaption}
\usepackage{bm}
\usepackage{xcolor}
\def\BibTeX{{\rm B\kern-.05em{\sc i\kern-.025em b}\kern-.08em
    T\kern-.1667em\lower.7ex\hbox{E}\kern-.125emX}}
\begin{document}

\renewcommand\citepunct{, }  
\renewcommand{\citedash}{--}  

\renewcommand{\vec}[1]{\boldsymbol{#1}}
\newcommand{\argmin}[1]{\underset{#1}{\mathrm{argmin}}}
\newcommand{\argmax}[1]{\underset{#1}{\mathrm{argmax}}}

\newcommand{\jul}[1]{{\color{blue} #1}}

\title{Stochastic Approximation of Variational Quantum Imaginary Time Evolution}

\author{
\IEEEauthorblockN{Julien Gacon\IEEEauthorrefmark{1}\IEEEauthorrefmark{2}, Christa Zoufal\IEEEauthorrefmark{1}, Giuseppe Carleo\IEEEauthorrefmark{2} and Stefan Woerner\IEEEauthorrefmark{1}
\\ \\}
\IEEEauthorblockA{\IEEEauthorrefmark{1}IBM Quantum, IBM Research Europe – Zurich, CH-8803 Rüschlikon, Switzerland} 
\IEEEauthorblockA{\IEEEauthorrefmark{2}Institute of Physics, École Polytechnique Fédérale de Lausanne (EPFL), CH-1015 Lausanne, Switzerland}
}

\maketitle

\begin{abstract}
The imaginary-time evolution of quantum states is integral to various fields, ranging from natural sciences to classical optimization or machine learning. 
Since simulating quantum imaginary-time evolution generally requires storing an exponentially large wave function, quantum computers are emerging as a promising platform for this task. 
However, variational approaches, suitable for near-term quantum computers, struggle with a prohibitive number of measurements and impractical runtimes for relevant system sizes.
Here, we suggest a stochastic approach to variational quantum imaginary-time evolution, which allows a significant reduction in runtimes. 
Our approach allows trading off invested resources and accuracy, which makes it also suitable for ground state preparation, where simulating the exact dynamics is not required.
We demonstrate the efficiency of our algorithm in simulations and show a hardware experiment performing the imaginary-time evolution of the transverse field Ising model on 27 qubits.
\end{abstract}


\section{Introduction}

Quantum imaginary-time evolution is a powerful tool which allows to prepare thermal states (or Gibbs states) and ground states of a quantum mechanical system~\cite{mcardle_variational_2019}.
Thermal states are particularly interesting in physics, for example, as they can be used to calculate thermodynamic observables~\cite{getelina_qmetts_2023,gacon_dual_2023}, or in machine learning, where they are used in quantum Boltzmann machines~\cite{zoufal_boltzmann_2021}. 
The preparation of ground states is even more general and finds applications in natural sciences, such as physics and chemistry, but also beyond quantum mechanical fields, such as classical optimization, finance, or machine learning~\cite{zoufal_blackbox_2023}.

Performing the imaginary-time evolution of quantum states generally requires storing an exponentially large wave function, and, therefore, quantum computers are emerging as a promising platform to solve this task. 
In contrast to real-time evolution, a unitary operation, imaginary-time evolution is non-unitary. Therefore, the standard Suzuki-Trotter approximation of the time-evolution operator~\cite{lloyd_universal_1996}, common in the real-time counterpart, cannot be applied directly to simulate imaginary-time dynamics on a gate-based quantum computer. 
Though there exist generalizations of Trotterization to imaginary-time evolution~\cite{motta_determining_2020}, these often require complex quantum circuits to be executed.
These requirements make Trotterization-based approaches unsuitable for near-term quantum computers, which are characterized by short qubit coherence times, limited connectivity, and noisy operations.
 
Instead of directly evolving the quantum state, variational approaches to imaginary-time evolution map the time evolution onto parameters in an ansatz circuit.
This ansatz circuit can be tailored to match the available device's capabilities
which makes variational methods especially prominent in the era of near-term quantum computers.
The mapping of state to parameter evolution in variational quantum imaginary-time evolution (VarQITE) can be achieved with a variational principle \cite{yuan_varqte_2019,mcardle_variational_2019}, such as the McLachlan variational principle, which relies on the evaluation of the Quantum Geometric Tensor (QGT) and the energy gradient at the current parameter values.

If the ansatz circuit has $d \in \mathbb{N}$ variational parameters, the calculation of the QGT requires sampling from $\mathcal{O}(d^2)$ circuits, respectively $\mathcal{O}(d)$ for the energy gradient \cite{stokes_qng_2020, gacon_qnspsa_2021}. 
This scaling is not an issue for variational states containing a small number of parameters. Still, it can quickly become a bottleneck on near-term devices for circuits with 100 or more parameters \cite{gacon_dual_2023}. 
As the current generation of quantum computers reaches 100 qubits and more, such as the IBM Quantum Eagle \cite{eagle} or Osprey \cite{osprey} devices, it is crucial to develop scalable algorithms that are suitable for the growing circuit sizes.

In this paper, we suggest a stochastic approach to variational quantum imaginary-time evolution (SA-QITE), that is based on stochastic approximation of Quantum Natural Gradients (QNG)~\cite{stokes_qng_2020, gacon_qnspsa_2021}.
Instead of computing the full QGT and gradient in each timestep, we start from an accurate initial estimation and correct the estimators in each iteration using unbiased samples. Unlike the $\mathcal{O}(d^2)$ scaling of the QGT, the samples rely on a simultaneous perturbation, stochastic approximation (SPSA) \cite{spall_2spsa_1997} method, which requires a constant number of circuits independent of the parameter dimension $d$ \cite{gacon_qnspsa_2021}.
We provide numerical evidence that SA-QITE requires fewer measurements than VarQITE to achieve the same accuracy and show that our algorithm represents a promising, near-term compatible imaginary-time simulation approach by applying it to a 27-qubit Ising model on an IBM Quantum processor.

Other approaches to avoid the evaluation of the QGT, that have recently been proposed rely on solving an optimization problem in each timestep, based on e.g. the fidelity \cite{gacon_dual_2023, benedetti_evolution_2021} or a purified Suzuki-Trotter step \cite{lin_qte_2021}. While these methods may exhibit a favorable scaling, it is challenging to measure the fidelity up to the required accuracy on current devices \cite{gacon_dual_2023}, or to efficiently measure the required state overlap \cite{benedetti_evolution_2021}.
Our stochastic approach does not suffer from these problems, as it relies only on relative differences of the state fidelity, and can readily be applied on near-term devices.

The remainder of this paper is structured as follows.
Section~\ref{sec:theory} starts by introducing VarQITE and then shows how to construct unbiased samples of the QGT and energy gradients and how to improve the estimator accuracy using momenta and exact initialization.
In Section~\ref{sec:numerical} we compare the resource requirements of SA-QITE and VarQITE in numerical simulations for imaginary-time evolution of the transverse field Ising model and, then, use a relaxed version of SA-QITE to solve a Max Cut optimization problem.
Next, we demonstrate our algorithm on a near-term quantum computer in Section~\ref{sec:hardware}, before concluding
in Section~\ref{sec:conclusion}.

\section{Stochastic Variational Imaginary Time Evolution}\label{sec:theory}

The normalized, imaginary-time evolution of an initial state $\ket{\Psi_0}$ under a Hamiltonian $H$ at time $t$ is defined as 
\begin{equation*}
    \ket{\Psi(t)} = \frac{e^{-tH}}{\sqrt{\braket{\Psi_0 | e^{-2tH} |\Psi_0}}} \ket{\Psi_0}.
\end{equation*}
In contrast to real-time evolution, which evolves under $\exp(-itH)$, the imaginary-time evolution operator $\exp(-tH)$ is not unitary.

Instead of evolving the quantum state directly, the idea of VarQITE is to project the state update to updates of variational parameters $\vec\theta \in \mathbb{R}^d$ in an ansatz state $\ket{\phi(\vec\theta(t))} \approx \ket{\Psi(t)}$.
This projection is achieved with a variational principle, such as McLachlan's variational principle, which allows computing the parameter derivative as the solution of the following linear system of equations
\begin{equation}\label{eq:mclachlan}
    g(\vec\theta)\dot{\vec\theta} = \vec b(\vec\theta),
\end{equation}
where we introduced the real part of the QGT $g = \mathrm{Re}(G) \in \mathbb{R}^{d\times d}$ and the evolution gradient $\vec b \in \mathbb{R}^d$.
The QGT is defined as 
\begin{equation}\label{eq:qgt}
    G_{ij}(\vec\theta) = \Braket{\frac{\partial\phi}{\partial\theta_i}| \frac{\partial\phi}{\partial \theta_j}} - 
     \Braket{\frac{\partial\phi}{\partial\theta_i} | \phi} 
     \Braket{\phi | \frac{\partial\phi}{\partial \theta_j}},
\end{equation}
and the evolution gradient is
\begin{equation}\label{eq:evograd}
    b_i(\vec\theta) = -\mathrm{Re}\left(\Braket{\frac{\partial\phi}{\partial \theta_i} | H | \phi}\right)
                    = -\frac{1}{2} \frac{\partial E}{\partial \theta_i},
\end{equation}
where $E(\vec\theta) = \braket{\phi(\vec\theta)|H|\phi(\vec\theta)}$ is the energy of the system. 
The individual terms of $g$ and $\vec b$ can be evaluated, for example, with a linear combination of unitaries approach (LCU) or parameter-shift rules \cite{schuld_gradients_2019}. These techniques require a constant number of expectation values per tensor element (or vector element) and therefore a total of $\mathcal{O}(d^2)$ circuits for the QGT, respectively $\mathcal{O}(d)$ for the evolution gradient.

Other variational formulations include the Dirac-Frenkel or the time-dependent variational principle, which also rely on the QGT and evolution gradient but may yield complex parameters \cite{yuan_varqte_2019}, which are not available in our quantum circuit model.

\subsection{Sampling the QGT and evolution gradient}

To circumvent the significant computational costs to evaluate the QGT in high-dimensional parameter spaces we 
replace $g$ with a stochastic estimate from which we can draw unbiased samples $\hat g$ at a constant cost \cite{gacon_qnspsa_2021}.
The samples are obtained by first reformulating the QGT as Hessian of the Fubini-Study metric and, then, estimating the Hessian 
using two nested simultaneous perturbation finite difference approximations, as
\begin{equation}\label{eq:g}
    \hat g = -\frac{1}{2}\frac{\delta F}{4 \epsilon^2}\frac{\vec\Delta_1 \vec\Delta_2^T + \vec\Delta_2 \vec\Delta_1^T}{2}
\end{equation}
where $\vec\Delta_{1,2} \sim \mathcal{U}(\{1, -1\}^d)$ are uniformly distributed perturbation directions, $\epsilon > 0$ is the perturbation magnitude, and
\begin{equation*}
    \begin{aligned}
    \delta F =\,&F(\vec\theta, \vec\theta + \epsilon(\vec\Delta_1 + \vec\Delta_2)) 
                - F(\vec\theta, \vec\theta + \epsilon(\vec\Delta_1 - \vec\Delta_2)) \\
                &- F(\vec\theta, \vec\theta + \epsilon(\vec\Delta_2 - \vec\Delta_1)) 
                + F(\vec\theta, \vec\theta - \epsilon(\vec\Delta_1 + \vec\Delta_2)),
    \end{aligned}
\end{equation*}
with the fidelity $F(\vec\theta, \vec\omega) = |\braket{\phi(\vec\theta)|\phi(\vec\omega)}|^2$.

There exist a variety of techniques to compute the fidelity $F$ of two quantum states prepared with quantum circuits $\ket{\phi(\vec\theta)}$ and $\ket{\phi(\vec\omega)}$, such as the Swap Test \cite{buhrman_swaptest_2001} and variations thereof \cite{cincio_learning_2018} or randomized measurements \cite{elben_overlap_2019}. Both these methods are, however, unsuitable for our near-term setting as the swap test requires doubling the circuit width and non-local operations, and the randomized measurements use an exponential number of measurements. Instead, we here use the compute-uncompute method \cite{havlicek_supervised_2019}, which prepares $U^\dagger(\vec\theta) U(\vec\omega)\ket{0}$ and estimates the probability of measuring $\ket{0}$. This doubles the circuit depth but does not add any additional qubits or couplings and is, therefore easier to execute on the near-term superconducting devices we consider.

Analogous to $g$ we can estimate $\vec b$ as a first-order gradient with a single perturbation direction $\vec\Delta \sim \mathcal{U}(\{1, -1\}^d)$,
\begin{equation}\label{eq:b}
   \hat{\vec b} = -\frac{1}{2}\frac{E(\vec\theta + \epsilon\vec\Delta) - E(\vec\theta - \epsilon\vec\Delta)}{2\epsilon} \vec\Delta.
\end{equation}

As we perturb all parameter dimensions at once, there is no dependency on the number of parameters $d$ and the calculation 
of a single sample $\hat g$ requires evaluating four circuits only and two expectation values for $\hat{\vec b}$.

\subsection{Improving estimator accuracy}

Since the samples $\hat g$ rely on only two perturbation directions (or one direction for $\hat{\vec b}$) 
they can have a very low accuracy. This is especially true for the QGT, since a single sample has at most rank 2, whereas the exact matrix can have a rank equal to the number of parameters $d$.
Therefore a single sample is typically replaced by an average over a batch of $N$ individual samples
\begin{equation*}
    \hat g_N = \frac{1}{N} \sum_{i=1}^N \hat g_{(i)}\text{ and }
    \hat{\vec b}_N = \frac{1}{N} \sum_{i=1}^N \hat{\vec b}_{(i)}.
\end{equation*}
The approximation error for both $G$ and $\vec b$ scales as $\mathcal{O}(N^{-1/2})$ in the number of samples
$N$, see also Appendix~\ref{app:qgt_error} for a numerical experiment.

The estimate at the current step can be combined with all previous ones to further increase stability. 
Refs.~\cite{gacon_qnspsa_2021, spall_2spsa_1997} suggest combining the samples from each time 
step into a global average
\begin{equation}\label{eq:qgt_global}
    \bar{g}^{(k)} = \frac{k}{k + 1} \bar{g}^{(k - 1)} + \frac{1}{k + 1} \hat{g}^{(k)}_N.
\end{equation}
For time evolution, however, a global average cannot correctly capture the time dependence of the QGT.
Instead, we propose to use momentum terms for both the QGT and the evolution gradient, such that the estimators in timestep $k$ are given by
\begin{equation*}
    \begin{aligned}
    \bar{g}^{(k)} &= \tau_1 \bar{g}^{(k - 1)} + (1 - \tau_1) \hat{g}_N^{(k)} \\
    \bar{\vec{b}}^{(k)} &= \tau_2 \bar{\vec{b}}^{(k - 1)} + (1 - \tau_2) \hat{\vec{b}}_N^{(k)},
    \end{aligned}
\end{equation*}
for momenta $\tau_1, \tau_2 \in (0, 1)$. 

As averaging by moment introduces a bias, especially at early times of the imaginary-time evolution, it is crucial to initialize the algorithm with accurate initial values of $g$ and $\vec b$.
These could be computed using resources that scale with $\mathcal{O}(d^2)$ a single time but can, in some cases, also be efficiently simulable classically.
For example, if the ansatz consists of Pauli rotations and CX gates and the initial parameters are integer multiples of $\pi/2$, every operation in the gradient calculations is a Clifford gate, as is discussed in detail in Appendix~\ref{app:clifford}. 
Two common scenarios where this is the case are classical optimization problems, which prepare an equal superposition state, $\ket{+}^{\otimes n}$, and use a QAOA or hardware-efficient ansatz \cite{farhi_qaoa_2014, zoufal_blackbox_2023}, or molecular ground state searches, where the ansatz is constructed from a Hartree-Fock initial state followed by operator evolutions, such as UCCSD, or partial swaps \cite{omalley_uccsd_2016, barkoutsos_ph_2018, romero_uccsd_2019}.

\subsection{Solving for the parameter update}

Determining the parameter derivative $\dot{\vec\theta}$ by directly solving the linear system in Eq.~\eqref{eq:mclachlan} is numerically only stable for the exact QGT and evolution gradient \cite{hackl_geometry_2020}. 
Here, however, we are dealing with a noisy linear system,
\begin{equation}\label{eq:noisy_lse}
    \bar g^{(k)}\dot{\vec\theta} = \bar{\vec b}^{(k)},
\end{equation}
due to a finite number of measurements in each circuit evaluation, a finite number of gradient samples $N$, and hardware noise.
These noise sources lead to an ill-conditioned linear system which requires careful regularization.

A simple regularization of the linear system is the addition of a weighted identity matrix to the system matrix, that is
\begin{equation}\label{eq:diagshift}
    (\bar g^{(k)} + \delta \mathbb{I})\dot{\vec\theta} = \bar{\vec b}^{(k)},
\end{equation}
for a shift $\delta > 0$ and the identity matrix $\mathbb I \in \mathbb{R}^{d\times d}$. This is equivalent to adding $\delta$ to each eigenvalue of the QGT estimate, decreasing the condition number, and improving the stability of the linear system.

Adding a diagonal shift, however, influences the parameter dynamics as the derivative magnitude is additionally restricted by its $\ell_2$ norm. In the case of optimization, for example, Ref.~\cite{gacon_qnspsa_2021} shows that QNG for large diagonal shifts approaches standard gradient descent.

To minimize the regularization effect on the evolution, we can solve for the update step only in a stable subspace, where the eigenvalues are above some threshold. Since the QGT estimate is real and symmetric, we can write
\begin{equation*}
        \bar g^{(k)}\dot{\vec\theta} = B\Lambda B^T \dot{\vec\theta} = \bar{\vec b}^{(k)}
\end{equation*}
for an orthonormal matrix $B$ and diagonal matrix $\Lambda = \mathrm{diag}(\lambda_1, \lambda_2, ..., \lambda_d)$, with the eigenvalues $\{\lambda_i\}_{i=1}^d$ of $\bar{g}^{(k)}$.
Defining $\dot{\vec\theta}^B = B^T \dot{\vec\theta}$ and $\vec b^B = B^T\bar{\vec b}^{(k)}$, we obtain the diagonal linear system
\begin{equation*}
    \Lambda\dot{\vec\theta}^B = \vec{b}^B,
\end{equation*}
which we solve by only considering well-conditioned components in the solution
\begin{equation}\label{eq:stable_subspace}
    \dot\theta^B_i = \begin{cases}
    b^B_i / \lambda_i, \text{ if } \lambda_i \geq \delta \\
    0, \text{ otherwise }
    \end{cases}.
\end{equation}
Finally we transform back to the original basis via $\dot{\vec\theta} = B\dot{\vec\theta}^B$.
This update rule ensures the parameter derivative does not diverge in ill-conditioned subspaces.
In Appendix~\ref{app:regularizations} we show that this approach produces a more stable convergence of SA-QITE.

\subsection{Relation to Quantum Natural Gradients}

Quantum Natural Gradient Descent (QNG) is a variant of gradient descent to find the minimum of an objective function $\ell(\vec\theta)$, where the size of the parameter update step is limited by the amount of change induced to the model, measured by the Fubini-Study metric \cite{amari_natural_1998, stokes_qng_2020}. 
The next step of the optimization, $\vec\theta^{(k+1)}$, is determined as 
\begin{equation*}
    \vec{\theta}^{(k+1)} = \argmin{\vec\theta \in \mathbb{R}^d} (\vec\theta - \vec\theta^{(k)})^T \nabla \ell(\vec\theta^{(k)}) + \frac{D^2(\vec\theta^{(k)}, \vec\theta)}{2\eta},
\end{equation*}
with the learning rate $\eta > 0$ and the Fubini-Study metric $D^2(\vec\theta, \vec\omega) = \arccos^2(|\braket{\phi(\vec\theta)|\phi(\vec\omega)}|)$.

By solving the minimization and approximating the Fubini-Study metric with a second-order Taylor expansion, we obtain the following update rule
\begin{equation*}
    \vec\theta^{(k+1)} = \vec\theta^{(k)} - \eta g^{-1}(\vec\theta^{(k)})\nabla\ell(\vec\theta^{(k)}),
\end{equation*}
which shows the analogy of QNG and VarQITE: If the QNG's loss function is $\ell(\vec\theta) = E(\vec\theta) / 2$ and we integrate the parameter in VarQITE using a forward Euler method with timestep $\eta$, the update rules of QNG and VarQITE coincide.

Therefore, the momentum-based estimations of the QGT and gradient are also suitable for optimization. 
In contrast to imaginary-time evolution, minimizing a loss function does not require tracking the parameter trajectory as closely as possible. This allows to relax the number of samples $N$ per iteration, which provides a less accurate estimation of $g$ and $\vec b$, but may decrease the total number of measurements required to converge to the minimum.

\section{Numerical Results}\label{sec:numerical}

In this section we investigate how SA-QITE performs for two tasks: the imaginary-time evolution of the transverse-field Ising model,
and the ground-state approximation of a diagonal Hamiltonian, typical e.g. in Max Cut problems.
All algorithms are implemented and simulated using Qiskit~\cite{qiskit}.

\subsection{Quantum Imaginary Time Evolution}

The transverse-field Ising model of $n$ spin-1/2 particles on a chain is given by
\begin{equation}
    H = J \sum_{i=1}^{n-1} Z_i Z_{i+1} + h \sum_{i=1}^n X_i,
\end{equation}
where we set the interaction to $J=1/2$, the transverse field strength to $h = -1$, $X_i$ and $Z_i$ are Pauli-$X$ and -$Z$ operators, acting on spin $i$.
As the initial state of the system, we consider $\ket{\Psi_0} = \ket{0}^{\otimes n}$.

\begin{figure}
    \centering
    \includegraphics[width=0.9\linewidth]{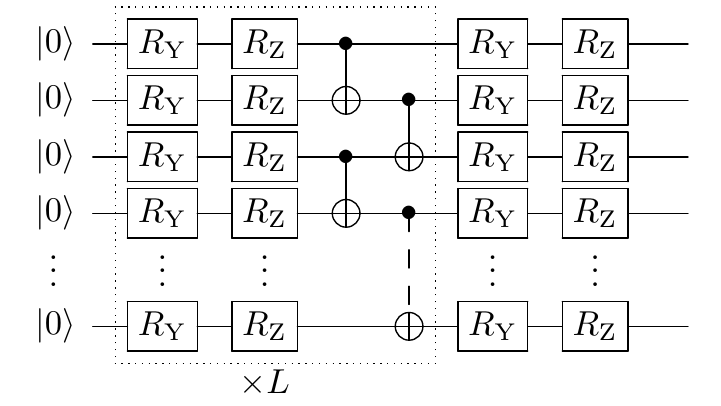}
    \caption{The structure of variational ansatz $\ket{\phi(\vec\theta)}$ used in the SA-QITE experiments.}
    \label{fig:efficientsu2}
\end{figure}

The variational ansatz $\ket{\phi(\vec\theta)}$ is chosen to reflect the nearest-neighbor connectivity of the Hamiltonian. It consists of $L$ alternating rotation layers, with parameterized Pauli-$Y$ and Pauli-$Z$ rotations, and entangling layers with pairwise CX connections. This circuit, whose structure is shown in Fig.~\ref{fig:efficientsu2}, has a CX depth of 2 per entangling layer and a total number of parameters $d = 2n (L + 1)$.

\begin{figure}[!t]
    \centering
    \includegraphics[width=\linewidth]{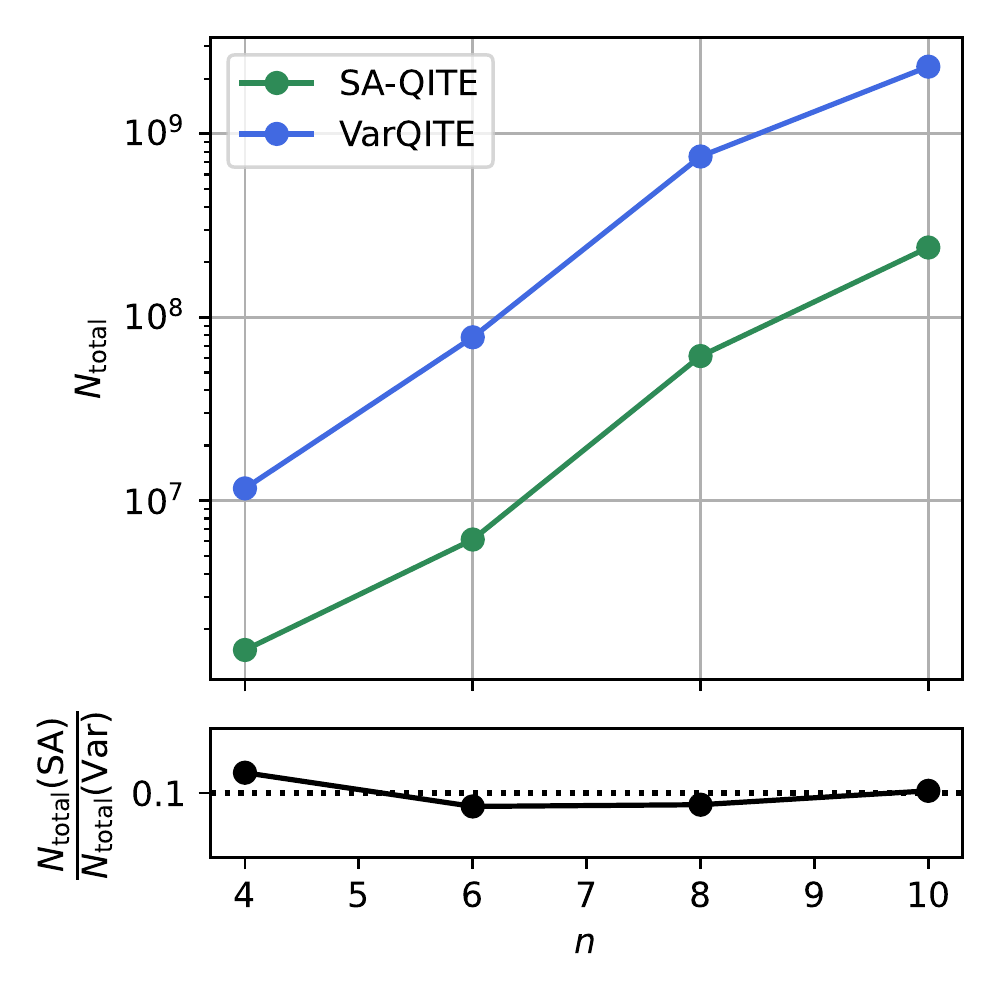}
    \caption{Total number of measurements, $N_\mathrm{total}$, required to achieve the target accuracy for SA-QITE and VarQITE, along with the fraction of both resource counts. SA-QITE requires $\approx 10\%$ of the number of measurements compared to VarQITE.}
    \label{fig:resources}
\end{figure}

We compare the total number of measurements required by SA-QITE and VarQITE to achieve an average integrated infidelity of $\mathcal{I} = 0.05$ over a time of $T=1.5$. Here, we define the infidelity with respect to the exact time-evolved state, $\ket{\psi(t)}$, that is 
\begin{equation}
    \mathcal{I}(T) = \frac{1}{T}\int_0^T \left(1 - |\braket{\phi(\vec\theta(t))|\psi(t)}|^2\right)\mathrm{d} t.
\end{equation}
In Fig.~\ref{fig:resources} we show the results for a varying number of qubits from $n=4$ to $10$, where we adjust the depth as $L=\lceil\ln{n}\rceil$.
The precise settings for each algorithm to achieve the target accuracy are listed in Appendix~\ref{app:benchmark}.

We observe that, on average, SA-QITE requires about one order of magnitude less measurements than VarQITE to achieve the target accuracy, while both algorithms exhibit the same asymptotic scaling.
Other variational time evolution algorithms based on optimizing a fidelity-based loss function, such as DualQITE \cite{gacon_dual_2023}, are also known to reduce the resources compared to VarQITE. However, since fidelity is difficult to measure to high accuracy on current devices, we here only focus on the stochastic approach.

\subsection{Ground state approximation}

\begin{figure}[!t]
    \centering
    \includegraphics[width=0.6\linewidth]{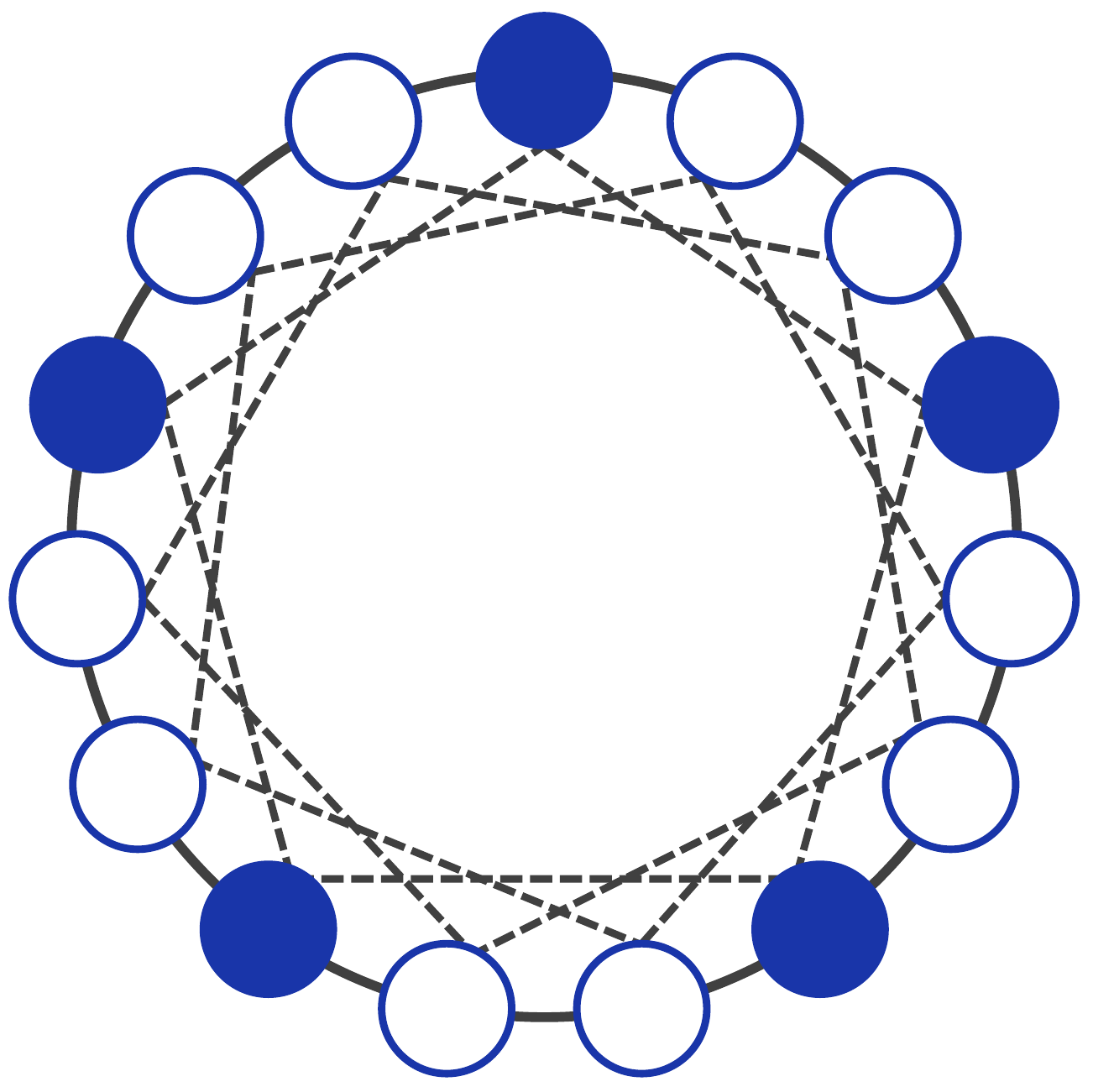}
    \caption{
        Solid lines mark interactions with $w_1 = 20$, dashed lines with $w_2 = -20$. 
        A optimal configuration is shown by filled and hollow circles, which indicate opposite qubit states.
        In total, there are 6 optimal configurations, which can be derived from the indicated solution by rotating the coloring (2 additional configurations) and inverting the colors (3 additional configurations).
    }
    \label{fig:circle}
\end{figure}

If the exact imaginary-time trajectory is not required, but we are only interested in approximating ground states of a Hamiltonian, we can relax the number of samples $N$ taken in each step.
As an example, we minimize the energy of a Hamiltonian derived from a Max Cut problem with integer weights on a circular graph with $n=15$ nodes, shown in Fig.~\ref{fig:circle}. The Hamiltonian is given by
\begin{equation*}
    H_C = w_1 \sum_{i=1}^{n} Z_i Z_{i+1 \text{ mod } n} + w_2 \sum_{i=1}^{n} Z_i Z_{(i + 3) \text{ mod } n},
\end{equation*}
with $w_1 = -w_2 = 20$.  

A widely-used approach to approximating the ground states of Hamiltonians obtained from a combinatorial optimization problem is using the 
Quantum Approximate Optimization Algorithm (QAOA)~\cite{farhi_qaoa_2014}.
There, the energy is minimized in a variational optimization with a specific ansatz that is motivated by simulated annealing from a
mixer Hamiltonian, $H_M$, whose ground state is easily prepared, to the target Hamiltonian, $H_C$.
The ansatz is defined as
\begin{equation*}
    \ket{\phi(\vec\gamma, \vec\beta)} = \left( \prod_{p=r}^{1} e^{-i\beta_p H_M} e^{-i \gamma_p H_C}\right) \ket{+}^{\otimes n},
\end{equation*}
with the mixer $H_M = -\sum_{i=1}^n X_i$, parameters $\vec\gamma, \vec\beta \in \mathbb{R}^{r}$, and we choose $r=2$. 
Since $H_C$ contains only two-qubit Pauli-$Z$ interactions, each term in the exponent $\exp(-i\gamma_p H_C)$ can 
be realized with a two-qubit Pauli rotation, $R_{ZZ}(2\gamma_p)$, on the interacting qubits. Similarly,
$\exp(-i\beta_p H_M)$ can be implemented with a layer of single-qubit $R_X(\beta_p)$ rotation gates.

\begin{figure}[!t]
    \centering
    \includegraphics[width=\linewidth]{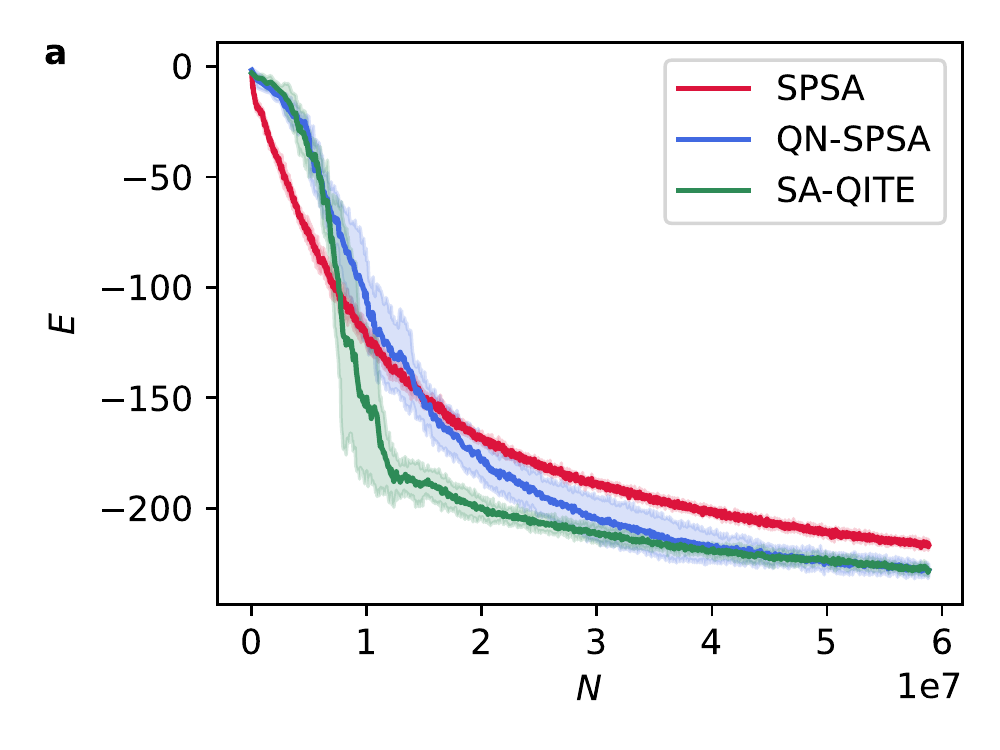}
    \includegraphics[width=\linewidth]{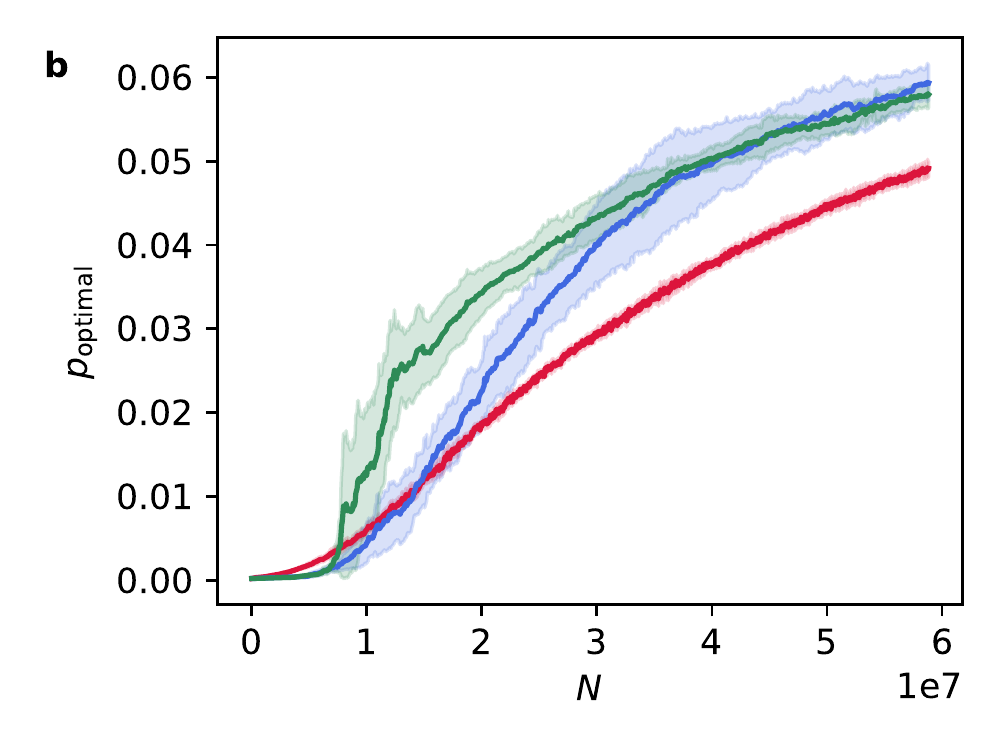}
    \caption{
    (a) The energies $E$ of the cost Hamiltonian as a function of total number of measurements $N$ for
    SPSA and the natural gradient adaptations.
    (b) The probability $p_\text{optimal}$ to sample one of the optimal states as a function of $N$.
    }
    \label{fig:qaoa_energies}
\end{figure}

The ground state of $H_M$ is $\ket{+}^{\otimes n}$, which is obtained by starting the optimization from
zero parameters, $\vec\beta = \vec\gamma = \vec 0$. In this case, the ansatz becomes a Clifford circuit 
and we can efficiently evaluate the QGT and energy gradients classically to initialize the SA-QITE algorithm.

We compare SA-QITE against SPSA, as gradient-based, measurement-efficient optimizer, and QN-SPSA, which this work is based on.
SPSA \cite{spall_2spsa_1997} minimizes the energy $E$ using unbiased gradient estimates, as in Eq.~\eqref{eq:b}, and for a learning rate $\eta > 0$ the update step is given by
\begin{equation*}
    \vec\theta^{(k+1)} = \vec\theta^{(k)} - \eta \frac{E(\vec\theta + \epsilon\vec\Delta) - E(\vec\theta - \epsilon\vec\Delta)}{2\epsilon}\vec\Delta,
\end{equation*}
where, as before, $\epsilon > 0$ is a small perturbation and $\vec\Delta \sim \mathcal{U}(\{1, -1\}^d)$ is the perturbation direction.
As a first-order gradient method, SPSA does not take into account the model sensitivity.
QN-SPSA \cite{gacon_qnspsa_2021} corresponds to SA-QITE without the improvements introduced in this paper, that is, QN-SPSA uses a global averaging of the QGT samples, as in Eq.~\eqref{eq:qgt_global}, and the identity as initial value, $g^{(0)} = \mathbb{I}$.

To ensure the optimizers converge towards the same minimum and avoid the saddle point at the zero initial point, we start the 
minimization from a small perturbation, $\vec\gamma^{(0)} = (10^{-3}, 10^{-3})$ and $\vec\beta^{(0)} = (10^{-2}, 10^{-2})$. 
Each circuit evaluation uses $8 \cdot 10^3$ measurements and we use a perturbation of $\epsilon = 10^{-2}$ for the gradient approximations.
For SPSA, we select the largest possible learning rate such that the algorithm converges consistently, which is the case for $\eta = 5 \cdot 10^{-7}$.
Since SA-QITE approximates QNG, the update step is normalized with respect to the induced change in the model and we can choose a substantially
larger learning rate, which translates to a ``timestep'' of $\Delta_t = 10^{-3}$.

As we are interested only in converging to the ground state, we relax the number of QGT and energy gradient samples to
$N_0=10$ in the first step, and reduce to $N_k = \max\{1, \lfloor (0.9)^{k} N_0 \rfloor\}$ in the $k$-th iteration.
The QGT momentum is set to $\tau_1=0.99$, but we use no gradient momentum ($\tau_2=0$) to avoid bias once the optimization converged.
Since we expect more noise in the QGT estimate as in the time-evolution case, we now solve the linear system with the diagonal shift of
Eq.~\eqref{eq:diagshift} with $\delta = 100$, which is roughly 0.5\% of the magnitude of the largest eigenvalue of the initial QGT.

In Fig.~\ref{fig:qaoa_energies}(a) we show the energy of the $H_C$ as a function of the total number of
measurements $N$. In the narrow loss landscape, SPSA, as gradient-based optimizer, is only able to converge slowly, even though it uses 
less resources per iteration. The natural gradient approximations QN-SPSA and SA-QITE, on the other hand, take into account the model sensitivity, which allows for a faster convergence. In particular we can see that, due to the accurate initial values of the QGT and energy gradient, SA-QITE initially performs better than QN-SPSA. Towards the minimum the QGT estimates of both these algorithms converge to the same values, which, in this example, leads to similar final energies.

Though the energy of the Hamiltonian is an indicator of solution quality, in a classical optimization problem we are often rather interested in the probability of sampling an optimal bitstring. In Fig.~\ref{fig:qaoa_energies}(b) we therefore show the probability $p_\text{optimal}$ to sample one of the optimal states, as described in Fig.~\ref{fig:circle}. 
The initial good performance of SA-QITE allows to amplify the solution probability most efficiently of the compared algorithms. To reach a 1\% overlap,
starting from the initial overlap of $\sim 2^{-15}$, for example, it requires only 64\% of the measurements of SPSA or QN-SPSA.

\section{Hardware Experiments}\label{sec:hardware}

\begin{figure}[!t]
    \centering
    \includegraphics[width=\linewidth]{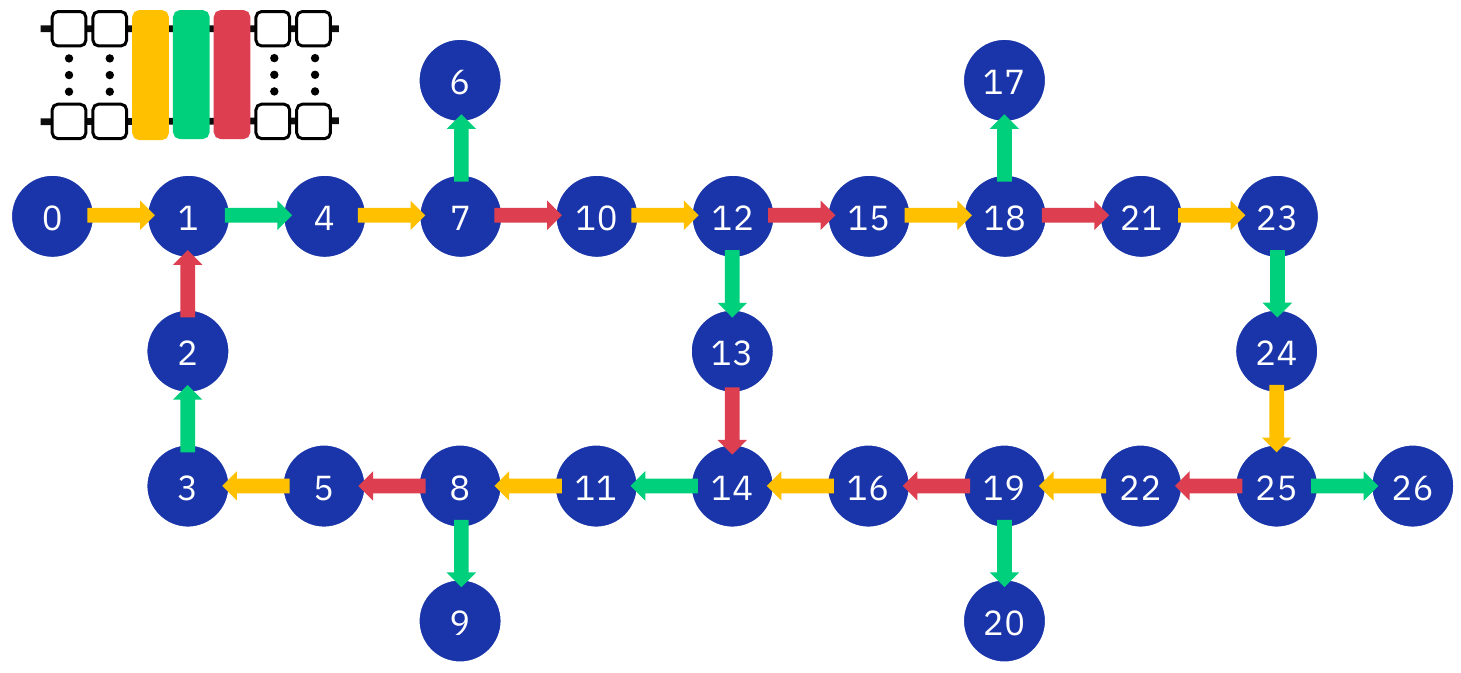}
    \caption{The coupling map of the 27-qubit quantum processor. The colors of the connections indicate the order in which the CX layers are implemented to achieve a CX depth of three. The arrows are pointing from control to target qubit.}
    \label{fig:falcon_chip}
\end{figure}

To test the near-term compatibility of SA-QITE, we scale the Ising Hamiltonian up to $n=27$ spins and execute the imaginary-time evolution on \texttt{ibm\_auckland}, which is one of the IBM Quantum Falcon processors \cite{ibm_quantum}.
Instead of spins on a chain, we consider nearest-neighbor interactions matching the topology of the device, shown in Fig.~\ref{fig:falcon_chip}. 
This allows us to choose an ansatz that is both hardware-efficient, as it has low depth when compiled to basic instructions of the quantum processor and problem-inspired, as it reflects the interactions of the Hamiltonian. 

The ansatz we use for this problem is similar to the one used in the numerical experiments, with the difference that we use a single entangling layer, $L=1$, and the pairwise connections exist between all qubit connections in the coupling map. This can be achieved with a CX depth of three.
Depending on the device, executing a large number of CX depth in parallel, or executing certain connection pairs at the time, can cause frequency collisions. In these cases, increasing the CX depth beyond the requirement minimum can still be favorable.

The parameters for the Ising Hamiltonian are set to $J=0.1, h=-1$ and we integrate up to $T=2$ with a timestep of $\Delta_t = 10^{-2}$. As before, the initial state is $\ket{0}^{\otimes n}$, which can be prepared with the variational ansatz by setting all initial parameter values to $0$. We perform SA-QITE with 1024 shots, $N=10$ samples per step, and momenta $\tau_1 = 0.99$ and $\tau_2 = 0$ and use no error mitigation. We use a Taylor expansion of the imaginary-time evolution operator as a classical reference calculation, as detailed in Appendix~\ref{app:27q_experiment}.

\begin{figure}
    \centering
    \includegraphics[width=\linewidth]{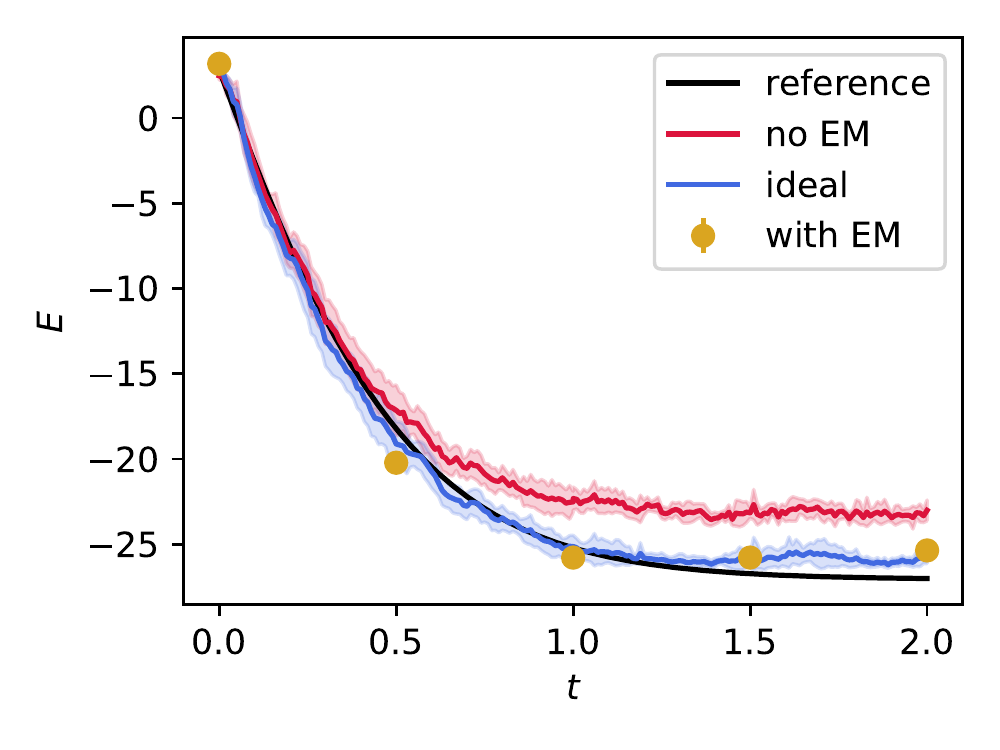}
    \caption{The energies for the imaginary-time evolution of the Ising model on 27 spins.
        While the non-error-mitigated values have an offset from the reference calculation,
        the exact evaluation of the parameter trajectory has a high accuracy, which is also reached by the error-mitigated points.
        }
    \label{fig:27q_energies}
\end{figure}

The results of the imaginary-time evolution are presented in Fig.~\ref{fig:27q_energies}. We see that, without error mitigation,
the energies calculated by SA-QITE follow the exact energies up to $t\approx 0.5$, but then reach a plateau at a constant offset of the reference calculation. 
However, if we evaluate the energies with the parameters obtained from the noisy hardware with an ideal statevector simulation instead, the energies are close to the exact solution.
This suggests that SA-QITE was able to find the right parameter trajectory despite the present hardware noise, even though the evaluated energies have some errors.

To test the quality of the noisy parameters, we use error mitigation to evaluate specific energies to a higher accuracy on another 27-qubit chip; \texttt{ibm\_peekskill} \cite{ibm_quantum}.
We mitigate readout errors with the matrix-free measurement mitigation (M3) \cite{nation_m3_2021} and employ a zero-noise extrapolation (ZNE) \cite{giurgica_zne_2020} to further account for errors during the circuit execution. Since ZNE is prone to coherent noise, we average each energy measurement over a set of twirled circuits, which reduces the noise into stochastic noise and thus improves the extrapolation performance \cite{kim_scalable_2023}. The workflow is summarized in Fig.~\ref{fig:em} and the techniques and specific settings are detailed in Appendix~\ref{app:27q_experiment}.

In Fig.~\ref{fig:27q_energies}, we see that error mitigation can improve the energy measurements significantly and approaches the ideal, statevector-based evaluation.

\begin{figure}
    \centering
    \includegraphics[width=\linewidth]{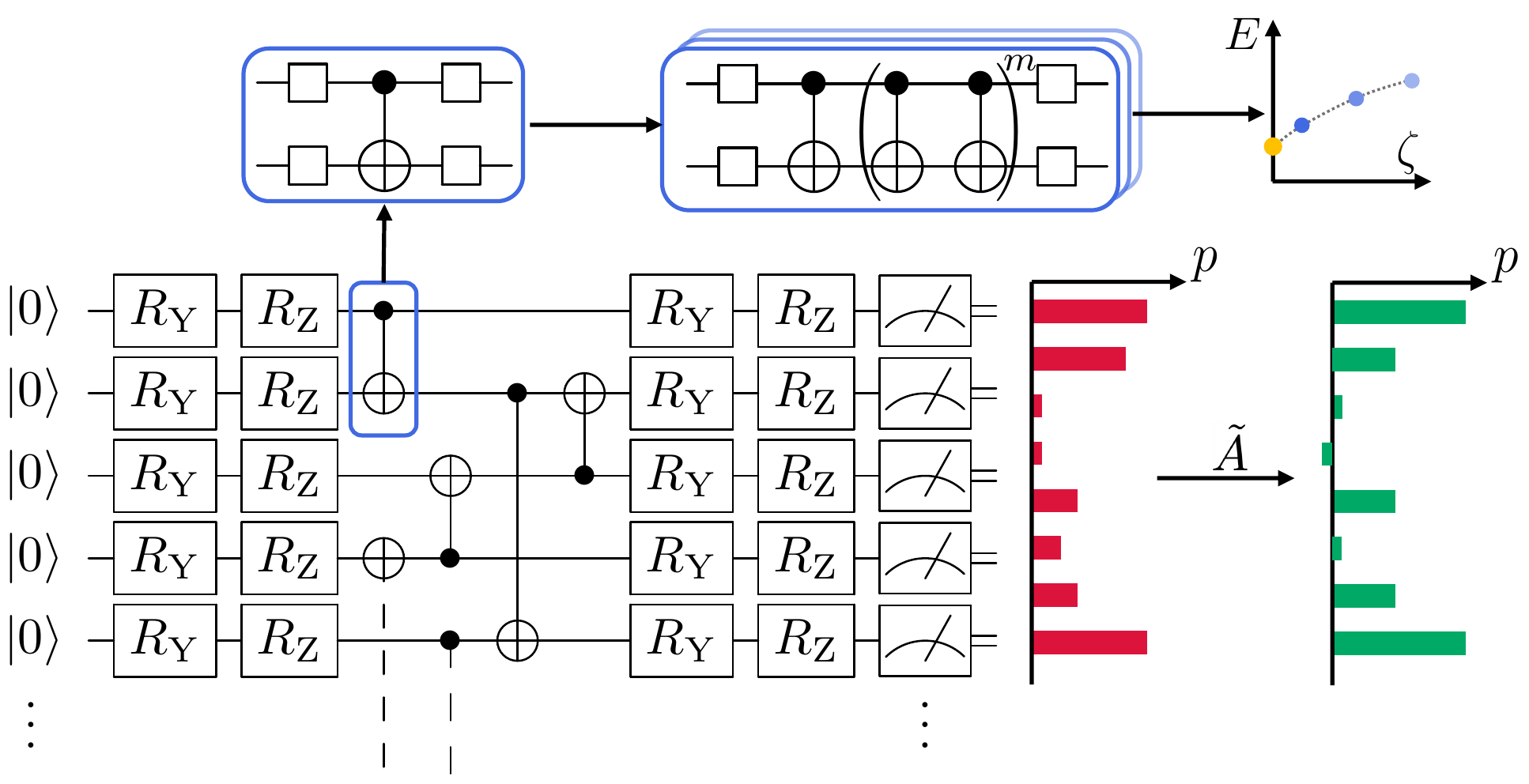}
    \caption{Error mitigation strategy for the energy measurements. Each CX gate is twirled using single-qubit Pauli operations before being folded $\zeta = 2m + 1$ times for $m \in \{0, 1, 2\}$.
             The energies of the folded circuits are readout mitigated using a reduced calibration matrix $\tilde{A}$. Finally, the energies are extrapolated to obtain the energy estimate $E(\zeta = 0)$.
            }
    \label{fig:em}
\end{figure}

\section{Conclusion}\label{sec:conclusion}

In this paper, we leverage a constant-cost sampling access to the QGT and energy gradient to implement a stochastic approximation of VarQITE, thereby reducing the prohibitive resource requirements.
The proposed SA-QITE algorithm for imaginary-time evolution is a generalization of the existing QN-SPSA \cite{gacon_qnspsa_2021} optimization algorithm, and uses a momentum-based combination of samples, combined an initialization with accurate initial values of the QGT and energy gradient. These changes are also applicable if SA-QITE is used for optimization instead of time evolution, in which case the number of samples per step could be relaxed.

In our numerical experiments we see that, compared to VarQITE, our SA-QITE algorithm reduces the number of measurements required to achieve a target accuracy by about one order of magnitude. For larger, overparameterized circuit models, where the QGT is costly to evaluate but does not have a complex structure and can be efficiently sampled, we expect the advantage of SA-QITE to further increase.
We have employed SA-QITE with a reduced number of samples for the optimization of a Max Cut Hamiltonian, where we showed that it is able to further improve on QN-SPSA, and is able to find optimal solutions at a lower number of measurements compared to other optimizers.

To demonstrate the near-term suitability of our algorithm, we use it to perform an imaginary-time evolution on a 27-qubit Ising model. There we find that even without error mitigation, the algorithm can determine the correct parameter dynamics. By investing additional resources for readout error mitigation and ZNE for individual points, we can retrieve energies close to the exact solution.

An open question for the family of stochastic variational algorithms of SA-QITE and QN-SPSA is how the number of samples could be chosen adaptively. This would allow for increasing the accuracy at times when the parameter dynamics are difficult to capture and using fewer resources if the samples have only a small impact on the current estimate. Another interesting question is how the method introduced here performs for adaptive circuit models, such as in ADAPT-VQE, where the number of parameters is small compared to other hardware-efficient models.

In conclusion, sampling from the QGT and energy gradients allows to reduce the number of measurements compared to VarQITE significantly, and is a suitable approach to implement imaginary-time evolution on near-term quantum computers. 
Scalable algorithms for current devices are one integral building block for solving practically relevant problems and pave the way to tackle remaining open questions, such as the construction of suitable variational circuit models.

\section*{Acknowledgments}

We thank Almudena Carrera Vazquez, Daniel Egger, Caroline Tornow, Max Rossmannek, Paul Nation and Christopher J. Wood for helpful discussions on the error mitigation techniques applied in this paper, and on the selection of the optimization problem. We are also grateful for the continuous support in the IBM Quantum stack provided by Jessie Yu, Kevin Tian, Daniel Kaulen, Diego Ristè, Maika Takita, and the whole IBM Quantum team.

We acknowledge the use of IBM Quantum services for this work. The views expressed are those of the authors, and do not reflect the official policy or position of IBM or the IBM Quantum team. 

IBM, the IBM logo, and ibm.com are trade marks of International Business Machines Corp., registered in many jurisdictions worldwide. Other product and service names might be trademarks of IBM or other companies. The current list of IBM trademarks is available at \texttt{https://www.ibm.com/legal/copytrade}.

\bibliographystyle{IEEEtran}
\bibliography{IEEEabrv,refs}

\appendices

\section{Sampling error for the QGT and evolution gradient}\label{app:qgt_error}

In this section, we investigate the convergence of the QGT and evolution gradient samples, $\hat g_N$ and $\hat{\vec b}_N$, as a function of the number of samples $N$. We measure the convergence as error in the $\ell_2$ norm, i.e.,
\begin{equation*}
    \|\hat g_N - g \|_2 = \sqrt{\sum_{i,j=1}^d \left( (\hat g_N)_{ij} - g_{ij} \right)^2},
\end{equation*}
and analogously for $\hat{\vec b}_N$.

\begin{figure}[htp]
    \centering
    \includegraphics[width=\linewidth]{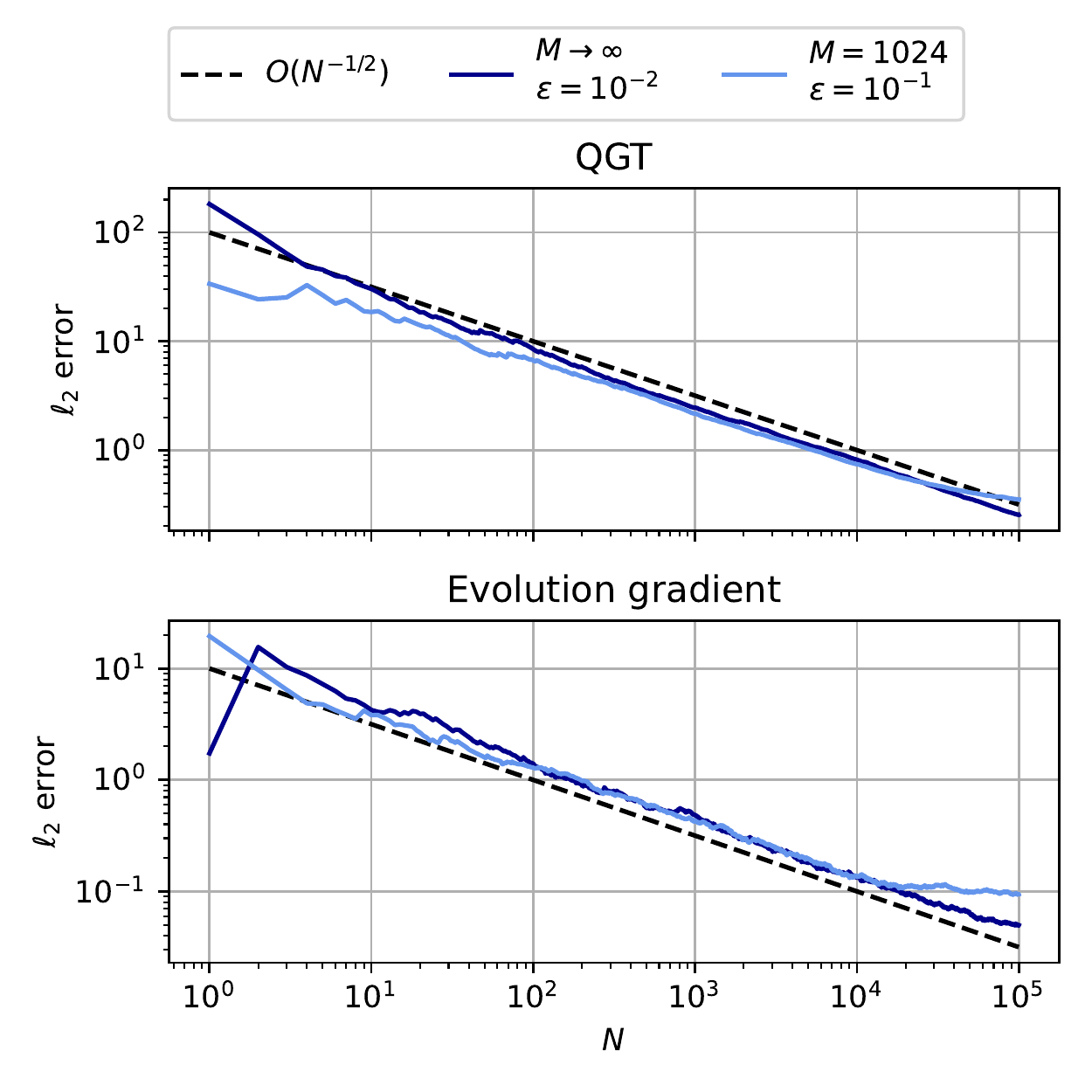}
    \caption{The sampling error of $\hat g_N$ and $\hat{\vec b}_N$ measured in $\ell_2$ distance to the exact values.
    The statevector-based evaluations ($M\rightarrow\infty$) use a perturbation of $\epsilon=10^{-2}$ and the measurement-based
    evaluations ($M=1024$) use $\epsilon=10^{-1}$.
    }
    \label{fig:graderrors}
\end{figure}

We calculate the errors for up to $N=10^5$ samples for the 8-qubit Ising Hamiltonian and variational ansatz used in the numerical experiments in Section~\ref{sec:numerical}, at the initial point $\vec\theta = \vec 0$. The errors are presented in Fig.~\ref{fig:graderrors}, for both a statevector-based evaluation of the circuits and for a shot-based evaluation with sampling statistics.
Both lines show the expected Monte Carlo convergence of $\mathcal{O}(N^{-1/2})$, however the finite readout accuracy in the shot-based case limits the achievable error. 
Therefore, increasing the number of samples beyond the shot-noise limit does not further improve the estimator accuracy.

\section{Clifford simulation of gradient circuits}\label{app:clifford}

If the ansatz circuit at the initial parameter values, $\ket{\phi(\vec\theta(0))}$, is a Clifford circuit, we can efficiently evaluate the its expectation values and gradients on a classical computer \cite{gottesman_stabilizer_1998}. This would allow the efficient evaluation of the initial QGT and evolution gradient, which we use in SA-QITE.

A large class of hardware-efficient circuits, including the ones used in this work, or problem-inspired circuits, such as the excitation-preserving or Trotterization circuits, are based on controlled Pauli gates and single-qubit Pauli rotations.
The controlled Paulis CX, CY, and CZ themselves belong to the Clifford group and the rotation gates become Clifford gates for certain rotation angles. For multiples of $\pi/2$, they can be expressed in terms of the Clifford gates $I,$ $S,$ $X$ and $H$:
\begin{equation*}
    \begin{aligned}
    R_X\left(\frac{\pi}{2}\right) &= S^\dagger H S^\dagger \\
    R_Y\left(\frac{\pi}{2}\right) &= X H \\
    R_Z\left(\frac{\pi}{2}\right) &= H R_X\left(\frac{\pi}{2}\right) H.
    \end{aligned}
\end{equation*}
Since $R_{X, Y, Z}(k\pi/2) = R^k_{X, Y, Z}(\pi/2)$, for $k\in\mathbb{Z}$ this shows the general case of any multiple of $\pi/2$.

To calculate the gradient 
\begin{equation*}
    \begin{aligned}
    \Ket{\frac{\partial}{\partial \theta_i} \phi(\vec\theta)} &= \frac{\partial}{\partial \theta_i} \prod_{k=d}^{1} U_k(\theta_k) \ket{0} \\
    &= U_d(\theta_d) \cdots \frac{\partial U_i}{\partial \theta_i}  \cdots U_1(\theta_1) \ket{0}
    \end{aligned}
\end{equation*}
we replace the unitary $U_i(\theta_i)$ with its derivative $\partial_i U_i(\theta_i)$.
Since, for the Pauli rotations, the derivative remains Clifford, the elements of the QGT and evolution gradient, defined in Eqs.~\eqref{eq:qgt} and~\eqref{eq:evograd}, can be calculated efficiently.

Other options to evaluate the gradients include using a parameter-shift rule with a shift of $\pi/2$ from the initial point,
or the linear-combination of unitaries method \cite{schuld_gradients_2019}, which only adds Clifford gates to the circuit.

\section{Regularization comparison}\label{app:regularizations}

In this section, we compare the different methods to solve the noisy linear system 
\begin{equation*}
    \bar g^{(k)} \dot{\vec\theta} = \bar{\vec b}^{(k)}.
\end{equation*}
We perform SA-QITE for the Ising Hamiltonian of Section~\ref{sec:numerical} for 8 qubits, using either the diagonal shift (Eq.~\ref{eq:diagshift}) or the stable-subspace (Eq.~\eqref{eq:stable_subspace}) method to solve the linear system in each timestep. For a numerically stable solution of the linear system with the diagonal shift, we rewrite the equation as convex, quadratic program, that is
\begin{equation*}
    \dot{\vec\theta} = \argmin{\vec x} \frac{\vec x^T (\bar g^{(k)} + \delta\mathbb I) \vec x}{2} - \vec x^T \vec b,
\end{equation*}
and solve it with the classical optimization routine COBYLA \cite{powell_cobyla_1994}.

\begin{figure}[htp]
    \centering
    \includegraphics[width=\linewidth]{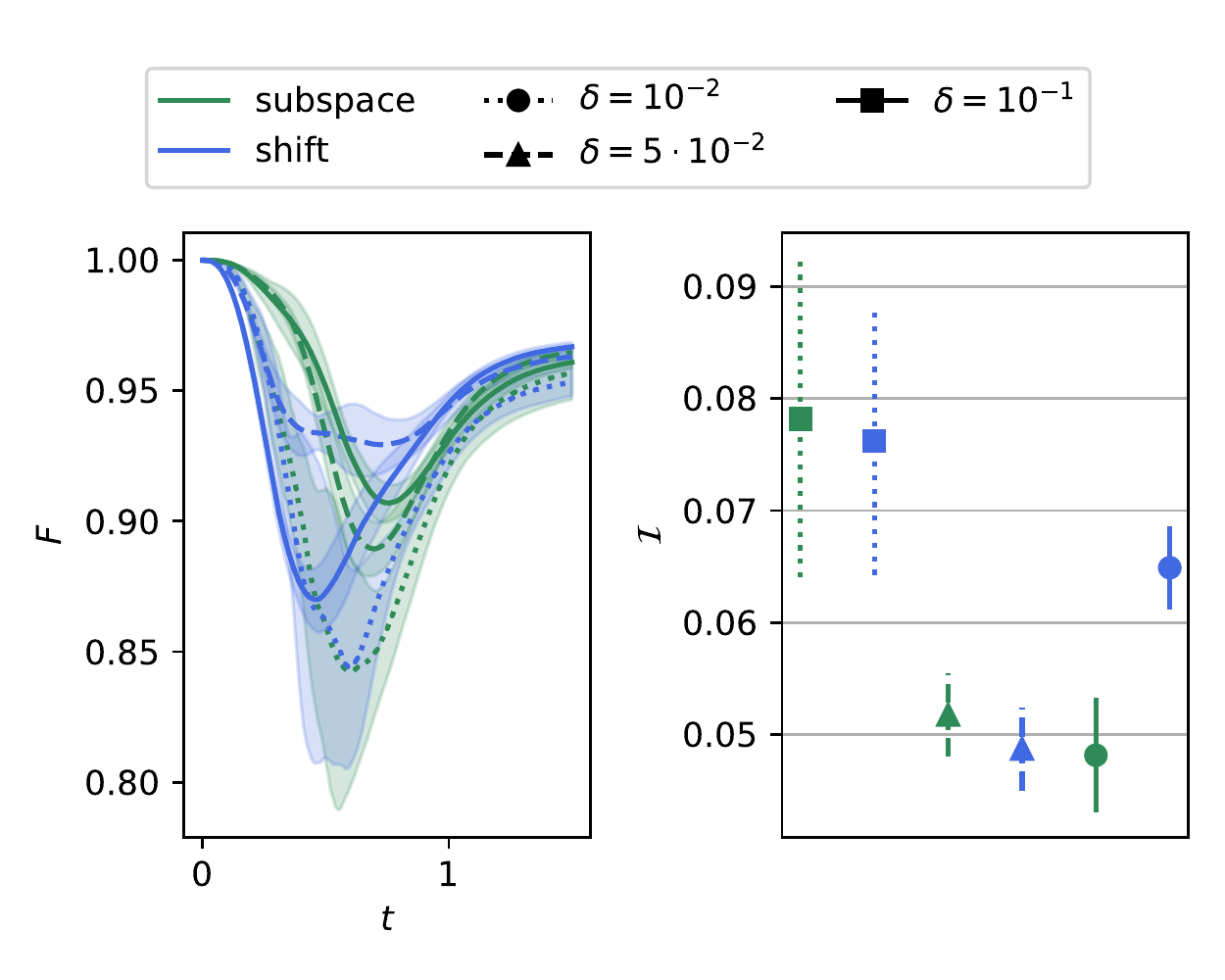}
    \caption{Fidelity $F$ compared to the exact time-evolved state at each time $t$, and integrated infidelity $\mathcal{I}$ for different linear system solvers and regularizations. In the stable subspace method (green), the regularization determines the eigenvalue cutoff threshold and in the diagonal shift technique (blue) the regularization equals the coefficient of the identity added to the QGT.}
    \label{fig:solvers}
\end{figure}

Figure~\ref{fig:solvers} shows the fidelity $F$ at each timestep $t$ and the integrated infidelity $\mathcal{I}$ for the two methods for different regularization constants $\delta$.
The stable-subspace technique provides the best results overall at $\delta=10^{-1}$. In addition, it seems to be the more stable because the integrated infidelity does not suffer from the same sudden increase at a larger regularization constant.
We, therefore, use the stable subspace technique for solving the linear systems in this work.

\section{Benchmark settings}\label{app:benchmark}

\begin{figure}[htp]
    \centering
    \includegraphics[width=\linewidth]{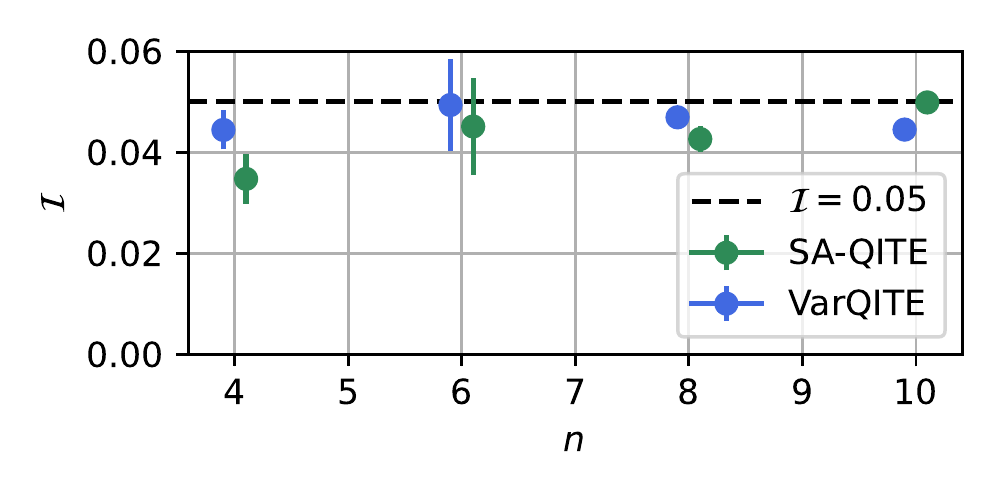}
    \caption{The mean and standard deviation of the integrated infidelity $\mathcal{I}$ for SA-QITE and VarQITE.}
    \label{fig:resource_accuracies}
\end{figure}

This section includes more details for the resource estimation in Section~\ref{sec:numerical}. In Fig.~\ref{fig:resource_accuracies} we show the achieved accuracies, measured in the integrated infidelity $\mathcal{I}$, for SA-QITE and VarQITE for the different numbers of qubits. The algorithm settings were each chosen, such that the accuracy is a close as possible to the threshold of $\mathcal{I} \leq 0.05$, and are displayed in Table~\ref{tab:settings}.

\begin{table}[htp]
    \centering
    \begin{subtable}{0.3\linewidth}
    \centering{
    \begin{tabular}{l|ll}
        $n$ & $M$ & $\delta$ \\ \hline
        4 & 128 & 0.05 \\
        6 & 400 & 0.05 \\
        8 & 1024 & 0.01 \\
        10 & 2048 & 0.05 
    \end{tabular}
    }
    \caption{VarQITE settings}
    \end{subtable}
    \begin{subtable}{0.6\linewidth}
    \centering{
    \begin{tabular}{l|lllll}
        $n$ & $M$ & $N$ & $\tau_1$ & $\tau_2$ & $\delta$ \\ \hline
        4 & 128 & 10 & 0.99 & 0.7 & 0.05 \\
        6 & 256 & 20 & 0.99 & 0.9 & 0.05  \\
        8 & 512 & 75 & 0.99 & 0.7 & 0.05  \\
        10 & 800 & 250 & 0.99 & 0.7 & 0.05  \\
    \end{tabular}
    }
    \caption{SA-QITE settings}
    \end{subtable}
    \caption{Algorithm settings for the resource estimations, including the number of qubits $n$, the number of measurements per basis $M$, the number of samples for the QGT and energy gradient $N$, the momenta $\tau_1$ and $\tau_2$, and the cutoff $\delta$ in the stable subspace solver.}
    \label{tab:settings}
\end{table}

\section{27-qubit experiment}\label{app:27q_experiment}

\begin{figure*}[!t]
    \centering
    \includegraphics[width=\textwidth]{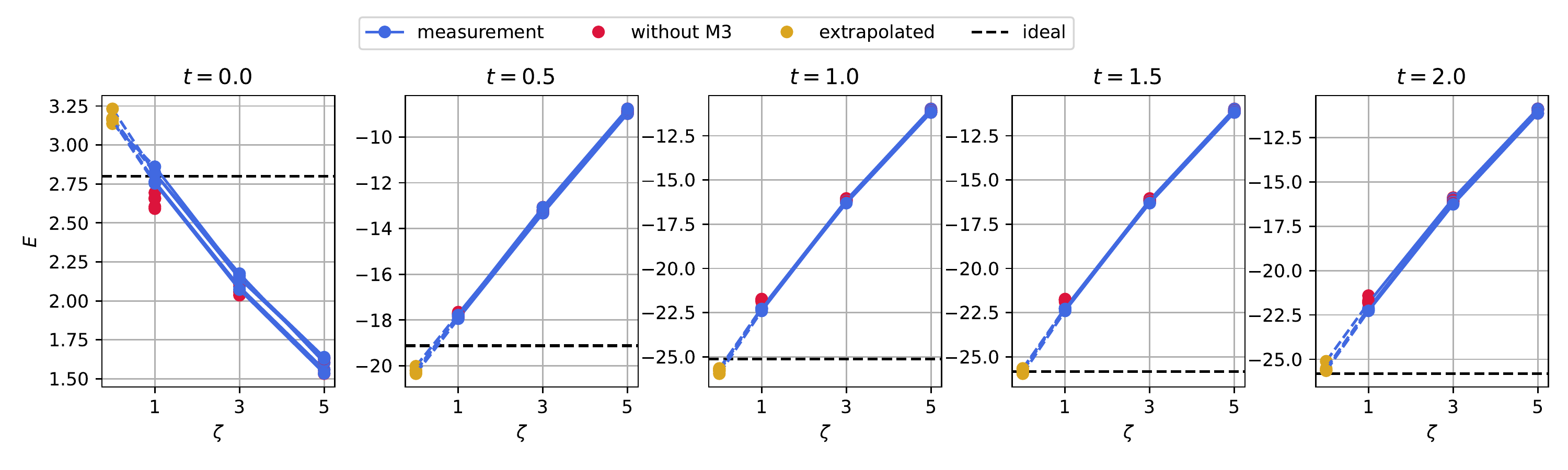}
    \caption{Zero-noise extrapolation for the energies at different times $t$. The figures show 5 independent repetitions of the extrapolation. Each interpolation point $E(\zeta)$ for $\zeta \in \{1, 3, 5\}$ is obtained by averaging over 25 Pauli Twirling instances and with M3 readout error mitigation. The black dashed line shows the ideal, statevector-based energy.}
    \label{fig:m3_zne}
\end{figure*}

\subsection{Classical reference solution}

For 27 qubits, computing the classical reference solution with a simple matrix exponential already requires too much memory to execute on an ordinary computer. Instead, to compute a classical reference solution, we Taylor-expand the imaginary-time evolution operator and normalize the state after each timestep. The update rule for the state $\ket{\psi(t)} \in \mathbb{C}^{2^n}$, is, then, given by
\begin{equation*}
    \begin{aligned}
        \ket{\tilde\psi(t + \Delta_t)} &= (\mathbb{I} - \Delta_t H)\ket{\psi(t)} \\
        \ket{\psi(t + \Delta_t)} &= \frac{\ket{\tilde\psi(t + \Delta_t)}}{\|\ket{\tilde\psi(t + \Delta_t)} \|_2},
    \end{aligned}
\end{equation*}
where $\mathbb{I}$ is the $2^n \times 2^n$ dimensional identity matrix.
We run the evolution for decreasing timesteps $\Delta_t$ and consider the reference solution as converged, if a smaller timestep does no longer changes the solution. In the 27-qubit experiment, this was reached for $\Delta_t = 10^{-3}$.

\subsection{Readout error mitigation}

Errors during the readout of qubit states are particularly dominating for shallow quantum circuits, like the circuit with a CX depth of three used in the hardware experiments in this work.
To mitigate these errors in a scalable approach, we use the M3 error mitigation \cite{nation_m3_2021}. In contrast to standard readout error mitigation approaches, such as a complete or tensored measurement mitigation, M3 corrects the measurements only in the subspace of measured bitstrings.

Assume the measured probability distribution over $n$ qubits and $N$ measurements is $\vec p$, where the $k$-th element describes the probability to measure the binary representation of $k$.
Then M3 computes the mitigated quasi-probability distribution $\vec q$ as 
\begin{equation}
    \vec q = \tilde{A}\vec p,
\end{equation}
with the truncated transfer matrix $\tilde A$. Note that $\vec q$ represents a quasi-probability distribution and can have negative vector elements, but still sums up to 1.
To construct $\tilde{A}$, we start from the the $2^n \times 2^n$-dimensional tensored transfer matrix $A = C_{n} \otimes \cdots \otimes C_1$, which is built from $n$ individual single-qubit calibration matrices of the form
\begin{equation}
    C_j = 
    \begin{pmatrix}
        p^{(j)}_{0 \rightarrow 0} & p^{(j)}_{0 \rightarrow 1} \\
        p^{(j)}_{1 \rightarrow 0} & p^{(j)}_{1 \rightarrow 1}
    \end{pmatrix},
\end{equation}
where $p^{(j)}_{a \rightarrow b}$ denotes the probability that qubit $j$ is initialized in state $\ket{a}$  and is measured to be in $\ket{b}$. 
The truncated matrix $\tilde{A}$ is obtained from $A$ by taking into account only indices, that are present in the noisy measurements $\vec p$, that is
\begin{equation}
    \tilde{A} = \left(A_{ij}\right)_{i,j \in \{k:~ p_k > 0\}}.
\end{equation}
Since the experiments in this paper used $N=1000$ shots for 27 qubits, $\tilde{A}$ has at most dimension $1000 \times 1000$, which is small enough to efficiently solve the linear system for the quasi-probabilities $\vec q$. For a larger number of shots, $\tilde{A}$ can be further truncated to include only transfers of bitstrings within a certain Hamming distance.
 
\subsection{Zero-noise extrapolation}

Zero-noise extrapolation (ZNE) is a generic error mitigation technique, which artificially amplifies noise sources to then extrapolate to the zero-noise limit. 
In this work, we apply ZNE to mitigate the errors introduced by two-qubit gates, in our case CX gates, as these are the main error sources during the quantum circuit execution.
We artifically amplify the CX noise by adding an additional even number of CX operations, which logically implement an identity but increase the error rate. 
Other amplifications are possible, such as the device-specific microwave pulse stretching, though here we focus on the identity-insertion as it is device agnostic and does not require additional pulse calibrations.

For a set of repetition levels $m \in \mathbb{N}$, we replace each single CX in the original circuit by 
\begin{equation*}
    \mathrm{CX} \rightarrow \mathrm{CX}^{2m + 1},
\end{equation*}
and denote the resulting measured energy as $E(\zeta)$, where $\zeta = 2m + 1$ is the number of applied CX gates.
As we expect the noise to increase exponentially in this gate repetition \cite{endo_zne_2018}, we then fit the measured energies for different $\zeta$ with 
an exponential model
\begin{equation*}
    E(\zeta; a, b, c) = a + b e^{c\zeta}, 
\end{equation*}
and extrapolate the ZNE estimate at $\zeta = 0$.

To reduce coherent errors in the CX applications, which can lead to unphysical extrapolations \cite{kim_scalable_2023}, we combine ZNE with Pauli Twirling.
Since on our the device, \texttt{ibm\_peekskill}, the single-qubit gate errors are much lower than the two-qubit gate errors, Pauli Twirling can be implemented by sandwiching each logical CX gate in between single-qubit Pauli operations, that preserve the logical operation.

In Fig.~\ref{fig:m3_zne} we show the interpolation energies $\{E(\zeta) : \zeta \in \{1, 3, 5\}\}$ and the extrapolated value at $\zeta = 0$ for states at different times $t$.
Each energy measurement includes M3 readout error mitigation and is averaged over $N_\text{twirl}=25$ Pauli Twirling instances, where each individual expectation value is measured with $N=1000$ shots per basis.

\end{document}